# Optical vector analysis with attometer resolution, 90-dB dynamic range and THz bandwidth

Ting Qing; Shupeng Li; Zhenzhou Tang; Bindong Gao; Shilong Pan*

Key Laboratory of Radar Imaging and Microwave Photonics, Ministry of Education, Nanjing University of Aeronautics and Astronautics, Nanjing, 210016, China

*Corresponding author: pans@nuaa.edu.cn

**Abstract:** Optical vector analysis (OVA) having the capability to achieve magnitude and phase responses is essential for fabrication and application of emerging optical devices. However, the conventional OVA often have to make compromises among resolution, dynamic range and bandwidth. Modulation-based OVA promises ultra-high resolution but suffers from measurement errors due to the modulation nonlinearity, narrow bandwidth restricted by the employment of microwave and optoelectronic devices, and difficulties in measuring optical devices with deep notch restricted by small signal modulation or the high-order sidebands. This paper proposes an original method to meet the measurement requirements for ultra-wide bandwidth, ultra-high resolution and ultra-large dynamic range simultaneously, based on asymmetric optical probe signal generator (ASG) and receiver (ASR). The use of the ASG and ASR not only doubles the measurement range without spectral aliasing, but also removes the measurement errors introduced by the modulation nonlinearity. The optimal signal modulation and enormous sideband suppression ratio in the ASG enables an ultra-large dynamic range. Thanks to the wavelength-independence of the ASG and ASR, the measurement range can boost to 2*N* times by applying an *N*-tone optical frequency comb (OFC) without complicated operation. In an experiment, OVA with a record resolution of 334 Hz (2.67 attometer in the 1550-nm band), a dynamic range of >90 dB and a measurement range of 1.025 THz is demonstrated.

 **Introduction**

Recently, optical devices to manipulate the magnitude and phase of optical signals with ultra-high resolution, ultra-wide bandwidth and ultra-large dynamic range are of fundamental importance for non-Hermitian photonics based on parity-time (PT) symmetry [1], optical nanoparticle detection [2], electromagnetically induced transparency [3], on-chip optical signal processing [4], ultra-sensitive optical sensing [5] and so on. These emerging optical devices put forward very urgent and stringent requirements of the measurement technology in terms of resolution, bandwidth, and dynamic range. For example, an ultra-narrow-band fiber Bragg grating (FBG) with a 3-MHz linewidth [6], an on-chip optical isolator with a bandwidth of 0.61 MHz [4], and a high-Q optical micro-resonator with a Q value of $1.7 \times 10^{10}$ (11.4-kHz or 91.2-attometer bandwidth in the 1550-nm band) [7] were proposed to improve the sensitivity of optical sensing system, which, obviously, needs an ultra-high resolution optical measurement method to implement the sensing demodulation. Similarly, ultra-narrow bandwidth phenomenon, such as spectral hole burning with a 172-kHz notch in Pr:YSO [8], PT-symmetry breaking with MHz-bandwidth resonance in whispering-gallery-mode microcavities [1], and ringing phenomenon in chaotic microcavity [5], has the capability of finely spectral manipulation, which also demands a measurement approach with attometer-level resolution. On the other hand, the spectra of a majority of optical devices and phenomena should be measured in a range of hundreds or even thousands of GHz. For instance, to measure the frequency response of the waveguide–ring resonator [9], the Fano resonance in all-dielectric metasurfaces [10], the accidental Dirac cone [11], and the hydrogen cyanide $H^{13}C^{14}N$ $2\nu_3$ [12], the bandwidth of the measurement systems should be in the order of THz. Besides, wide bandwidth and high-resolution measurement are simultaneously needed in some systems [13]. Dynamic range is another important parameter to evaluate a measurement method and concerned by numerous applications. For example, an FBG is expected to have a 50-dB extinction ratio, but the measured result is limited to about 35 dB due to the low dynamic range of the measurement apparatus [14].

Although researchers have been striving to achieve these goals for decades and some optical vector analyzers were previously reported to obtain the magnitude and phase responses [15-30], few can perform the measurement with simultaneous ultra-high resolution, ultra-large dynamic range and ultra-wide bandwidth. Table 1 shows a comparison of different kinds of optical vector analysis (OVA). As can be seen, the OVA based on the interferometry method [15] can provide a wide measurement range and a large dynamic range, but a poor resolution. Optical channel estimation (OCE) [16] can reach a sub-MHz resolution, but it is vulnerable in the dynamic range and measurement range. The OVA based on optical single-sideband (OSSB) modulation theoretically has the potential of reaching a sub-Hz resolution [17-27], but the existence of the high-order sidebands will severely degrade the resolution[23], introduces considerable measurement errors and restricts the dynamic range [22,

23]. In addition, the bandwidth of the electro-optical conversion devices or microwave components usually limits the measurement range to tens of GHz.

Table 1. Typical performance of the optical vector analyzers

| Parameter | Interferometry [15] | Optical channel estimation [16] | OSSB-based OVA | Proposed OVA |
|---|---|---|---|---|
| Resolution | 200 MHz | 1.25 MHz | 23.4 kHz [27] | 334 Hz |
| Measurement range | Several THz | 10 GHz | 105 GHz [24] | 1.025 THz |
| Dynamic range | 60 dB | 25 dB | 60 dB | >90 dB |

In this article, we propose and demonstrate, for the first time to the best of our knowledge, a novel method to perform optical vector analysis with simultaneous ultra-high resolution, ultra-large dynamic range and ultra-wide bandwidth. The basic idea is to generate asymmetric optical probe signal for carrying on the magnitude and phase responses of the optical device under test (ODUT), and to detect the responses without spectral aliasing by an asymmetric signal receiver (ASR). The use of the asymmetric optical probe signal generator (ASG) and ASR in the OVA yields three prominent benefits. First, the high-order sidebands inevitably generated by the modulation nonlinearity have a neglected influence on the measurement resolution, because the undesirable components introduced by the high-order sidebands have different frequencies with the useful ones in the ASR, removing the primary source of the measurement error in the conventional modulation-based OVAs. Second, thanks to the extraordinarily high-efficiency frequency-shifting of the acoustooptic modulator (AOM) in the ASG, the unwanted residual sideband [28] is nearly absent, leading to an enormous sideband suppression ratio (SSR). By applying an optimal modulation index, the proposed method can achieve an ultra-large dynamic range. Finally, thanks to the wavelength-independence of the ASG and ASR, any comb line from an optical frequency comb (OFC) can be selected as the optical carrier for a wavelength channel. Thus a large measurement range can be realized by stitching several consecutive channels. Besides, the measurement range in each channel is doubled by using the asymmetric optical probe signal, so when using an $N$-tone OFC the measurement range can boost to $2N$ times without complicated operation.

**Results**

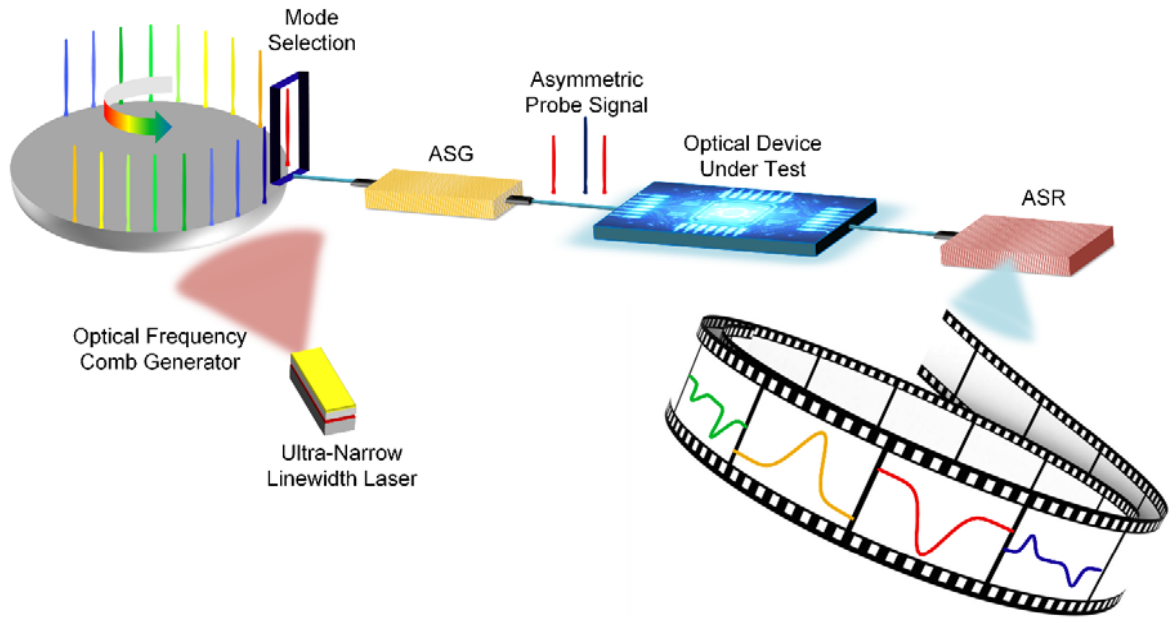

**Fig. 1** Principle of the optical vector analysis using an optical frequency comb. ASG, asymmetric optical probe signal generator; ASR, asymmetric optical probe signal receiver. The mode selection module selects a comb line from the OFC signal to generate an asymmetric probe signal by the ASG. After the ODUT, the information carried by the asymmetric probe signal is extracted by the ASR. Changing the wavelength of the mode selection module, responses in all wavelength channels can be obtained.

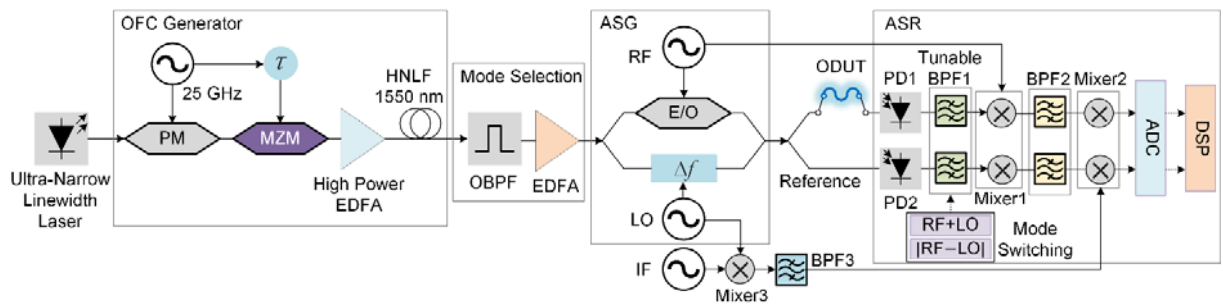

**Fig. 2** Experimental setup of the proposed OVA. OFC, optical frequency comb; PM, phase modulator; MZM, Mach-Zehnder modulator; OBPF, optical bandpass filter; EDFA, erbium-doped fiber amplifier; HNLF, highly nonlinear fiber; ASG, asymmetric optical probe signal generator; RF, radio frequency; LO, local oscillator; E/O, electro-optic; ODUT, optical device under test; ASR,

asymmetric optical probe signal receiver; PD, photodetector; BPF, optical bandpass filter; IF, intermediate frequency; ADC, analog-to-digital converter; DSP, digital signal processor.

**Principle and experimental setup**

Figure 1 illustrates the conceptual diagram of the OVA. The optical signal from an ultra-narrow linewidth laser goes through an OFC generator to stimulate an ultra-narrow linewidth OFC signal. A mode selection module is used to select a comb line, which is sent to an ASG to generate an asymmetric optical probe signal. The asymmetric optical probe signal propagates in the optical device under test and carries on the optical responses in both sides of the comb line. An ASR extracts the spectral response from the asymmetric optical signal, removing the influence of the high-order sidebands and other unwanted components. By sweeping the frequency of the asymmetric optical probe signal via tuning an RF source, the responses in the wavelength channel corresponding to the selected comb line would be obtained. Selecting other comb lines to perform the measurement and stitching the measured responses in all wavelength channels leads to the measurement of wideband responses covered by the OFC.

Figure 2 illustrates the experimental setup of the proposed OVA, which consists of an ultra-narrow linewidth laser (OEwaves OE4010) with a linewidth of 300 Hz, an OFC generator, a mode-selection module, an ASG and an ASR. The OFC generator is comprised of a phase modulator (PM, EOSPACE), a Mach-Zehnder modulator (MZM, Fujitsu FTM7938EZ), a high power erbium-doped fiber amplifier (EDFA, Amonics AEDFA-33-B-FA) and a highly nonlinear fiber (HNLF, 1550 nm), which generates a 41-tone OFC with a fixed frequency spacing. A tunable optical bandpass filter (TOBPF, WaveShaper 4000s) is followed to select the $n$th ($1 \leqslant n \leqslant 41$) comb line from the OFC and another erbium-doped fiber amplifier (EDFA, Amonics AEDFA-35-B-FA) is inserted to amplify it. In the ASG, the selected comb line is divided into two portions. One portion goes through an AOM (Gooch&Housego), which plays as an effective frequency shifter to have the frequency of the signal upshifted by a frequency of $\Delta\omega$=80 MHz. Thus, a frequency-shifted optical carrier is obtained. The other part of the optical signal is modulated by a frequency-sweeping RF signal (denoted as $\omega_e$) at another MZM (Fujitsu FTM7938EZ) to generate a carrier-suppressed optical double-sideband (ODSB) modulation signal consisting of two sweeping ±1st-order sidebands. Then, the two signals are combined to form an asymmetric optical probe signal. The asymmetric optical probe signal is further divided into two paths and finally sent to the ASR. The ASR consists of two photodetectors (PDs, U²T 2120RA) and an electrical signal processing module. In the upper path (measurement path), the transmission response of the ODUT modifies the magnitude and phase of the asymmetric optical probe signal. After the square-law detection in PD1, two RF components (denoted as $S_{mea}$) carrying the information of the ODUT are generated at the frequencies of $|\omega_e-\Delta\omega|$ and $\omega_e+\Delta\omega$ by beating the ±1st-order sidebands and the frequency-shifted carrier. It is worth noting that the existence of the high-order sidebands has no influence on the measurement results due to the difference in frequency except some predictable and removable points where $|\omega_e\pm\Delta\omega|=n\omega_e$ or $|\omega_e\pm\Delta\omega|=|n\omega_e\pm\Delta\omega|$. In the lower path (reference path), the asymmetric optical probe signal is directly sent to another PD (PD2 in Fig. 2), such that two reference RF signals (denoted as $S_{ref}$) with the frequencies of $\omega_e+\Delta\omega$ and $|\omega_e-\Delta\omega|$ without the phase and magnitude changes introduced by the ODUT are generated. A wideband tunable electronic filter containing a series of electrical switches and parallel tunable bandpass filters (BPFs) is used to select the component of either $\omega_e+\Delta\omega$ or $|\omega_e-\Delta\omega|$. Mixer1 converts the $\omega_e+\Delta\omega$ and $|\omega_e-\Delta\omega|$ components into $\Delta\omega$, BPF2 suppresses the unwanted components generated by the Mixer1, and then Mixer2 converts the $\Delta\omega$ component into an intermediate frequency (IF). The IF receiver would have a larger dynamic range than a baseband receiver because the latter would suffer from quadrature phase errors and power imbalance of IQ-demodulators. To sample the measurement signal and the reference signal, ADCs with large effective number of bits are used. Measurement results are achieved in a DSP via $S_0=S_{mea}\div S_{ref}$, which can remove and the common-mode noise in the measurement and reference paths. To further increase the accuracy of the measurement, the system response and the difference between the measurement and reference paths should be taken into account. To do so, we can remove the ODUT and directly connect the two test ports. In that case, another measured response $S'_{mea}$ and its reference $S'_{ref}$ will be obtained. We then get a calibration parameter $S_{cal}=S'_{mea}\div S'_{ref}$, and an accurate response can be achieved via $S=S_0\div S_{cal}$ (see Methods). In our implementation, the ADCs, DSP, BPF2, and Mixer2 are realized by a receiver of an electrical vector network analyzer (EVNA, R&S ZVA67). The incoherent optical structure and coherent electrical receiver make the ASR more sensitive to detect weak signal and more stable than the interferometry-based method [15].

**Implementation of ultra-wide bandwidth**

An experiment based on the setup shown in Fig. 2 is performed. An OFC signal with a frequency spacing of 25 GHz is generated, and 41 comb lines with relatively high power are selected to be the optical carriers. Figure 3 depicts the optical spectrum of the OFC and the optical carriers selected from the OFC by the mode selection module. As can be seen in Figs. 3(b) and 3(c), the side-mode suppression ratio (SMSR) of the selected comb line is 42.07 dB in the middle of the OFC and 21.22 dB in the marginal area. Sweeping the frequency of the RF signal to cover the half of the frequency spacing of the OFC, i.e., 12.5 GHz, the -1st- and +1st-order sideband would probe the full frequency response on both sides of the selected comb line in the range of 25 GHz (one wavelength channel).

Fig. 4 shows the frequency responses of a sampled FBG (red lines), which is measured by the proposed OVA. Assume the frequency of the selected comb line is $\omega_o$, the measurement of the wavelength channel is performed by three consecutive segments [$\omega_o$−12.5 GHz, $\omega_o$−80 MHz], [$\omega_o$−80 MHz, $\omega_o$] and [$\omega_o$, $\omega_o$+12.5 GHz], corresponding to the beat notes with frequencies of $\omega_e$−$\Delta\omega$ ($\omega_e$>$\Delta\omega$), $\Delta\omega$−$\omega_e$ ($\omega_e$<$\Delta\omega$) and $\omega_e$+$\Delta\omega$, respectively. Thanks to the continuity of the responses of the sampled FBG in any two adjacent channels, the measured responses in different channels can be stitched together regardless of the power difference of the comb lines. In addition, because of the excellent frequency stability of the OFC, the proposed method can extend the measurement range tremendously and the measurement resolution will not be deteriorated. In the experiment, all the 41 comb lines with a frequency spacing of 25 GHz are selected one by one as the optical carrier, which boosts the measurement range to 1.025 THz. As a comparison, an amplified spontaneous emission (ASE, Amonics AEDFA-35-B-FA) source and a 0.1-pm resolution optical spectrum analyzer (OSA, BOSA 400) are used to measure the magnitude response of the ODUT, with the result shown as the black line in Fig. 4(a). As can be seen, the two measurements agree very well. It should be noted that the signal to noise ratio (SNR) of the measured responses in the marginal areas is lower than that in the middle, because the comb lines in those areas have smaller power. The measurement range of the proposed OVA mainly depends on the coverage area of the selected optical comb lines. If the frequency spacing of the OFC and the number of the comb lines are enlarged [31], the measurement range can be extended to tens or even hundreds of THz in theory.

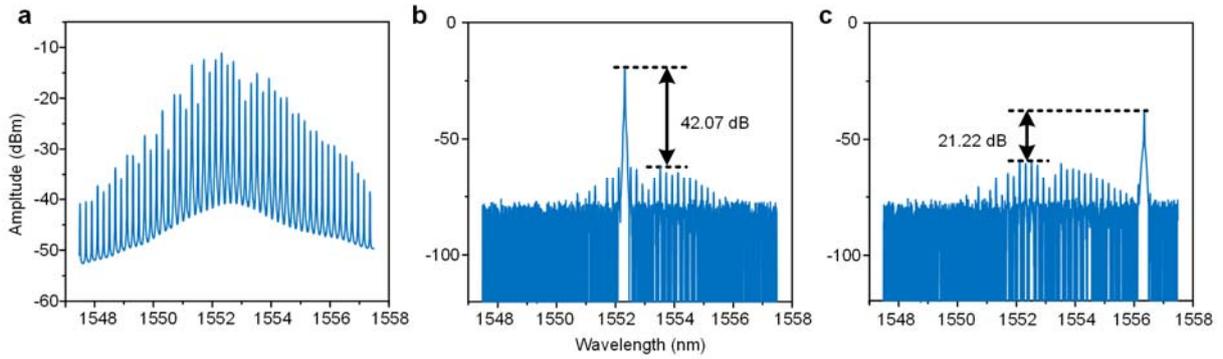

**Fig. 3** The optical spectra of the OFC and the selected comb line. **a** The generated OFC signal is measured in a span of 10 nm, and the comb lines in the middle range of 8 nm would be used in the experiment. **b** The comb line in the middle of the OFC selected by a TOBPF which has an SMSR of 42.07 dB. **c** The comb line with an SMSR of 21.22 dB in the marginal area of the OFC selected by a TOBPF.

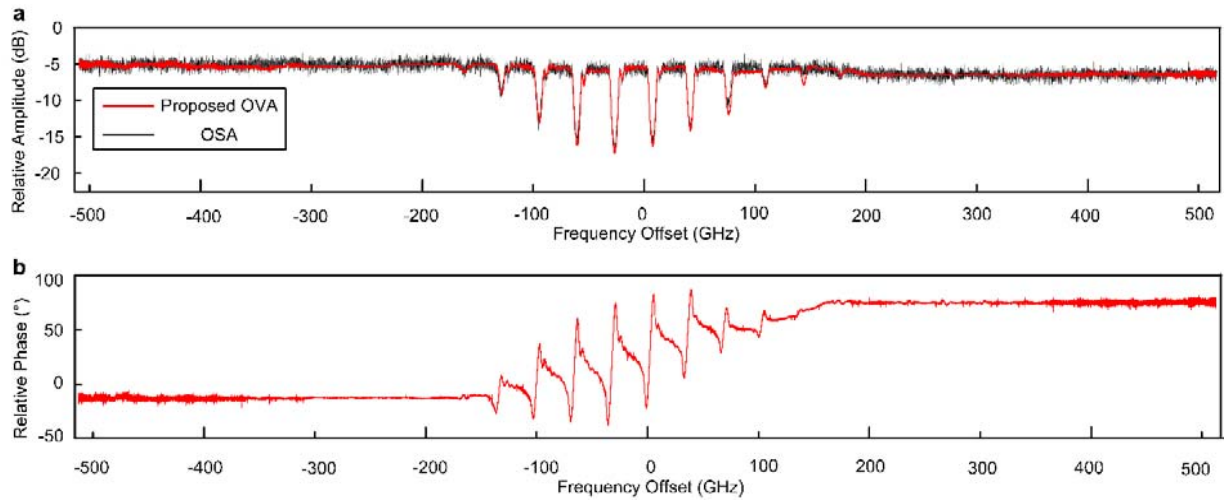

**Fig. 4** The spectral responses of a sampled FBG measured by the proposed OVA. **a** The magnitude responses are measured by the proposed method (red line) and an OSA (black line). **b** Since the OSA cannot measure the phase of the optical signal, the phase response can only be obtained by the proposed OVA.

**Implementation of ultra-high resolution**

For the conventional OSSB-based OVA, the existence of the high-order sidebands greatly degrades the wavelength resolution [23]. This resolution limitation factor is fully removed by the proposed method, because the frequencies of the beat notes between the high-order sidebands are $m\omega_e$ ($m$=1,

2,...) and the beating between the high-order sidebands and the frequency-shifted optical carrier generates $|n\omega_e\pm\Delta\omega|$ ($n\geq2$) components, which are different from the required $|\omega_e\pm\Delta\omega|$ components and therefore automatically dropped by the coherent electrical receiver in the ASR. With the influence of the high-order sidebands eliminated, the resolution of the proposed OVA only depends on the frequency step of the frequency-swept RF source and the linewidth of the laser. In the experiment, the frequency step of the frequency-swept RF source can reach a few sub-Hz and the laser has a linewidth of 300 Hz, so the proposed method can realize an ultra-high resolution. In addition, the existence of the reference path can remove the common noise or phase jitter between the measurement path and the reference path introduced by the optical source and the environment, so we can eliminate the time-varying measurement errors, ensuring high measurement accuracy and resolution.

To verify the ultra-high resolution, an ultra-narrow-bandwidth phase-shifted FBG is measured by the proposed OVA. Figs. 5(a) and (b) show the frequency responses in the measurement and reference paths, respectively. The final measurement results achieved after calibration are shown in Fig. 5(c). As can be seen, time-varying measurement errors are well suppressed. Fig. 5(d) shows the zoom-in view of the zone of interest, in which the frequency range is 20 MHz and the resolution is 334 Hz. Since there is no other existing approach to achieve this resolution, we verify the measurement results through the Kramers-Kronig relationship (KKR) between the phase and magnitude responses. The phase response calculated from the magnitude response via the KKR is shown as the black line in Fig. 5(d). As can be seen, the calculated result fits the measured phase response very well, which proves that the proposed method can provide precise measurement in such high resolution. It should be noted that if the linewidth of the laser can be pushed to several Hz, the resolution of the proposed method might reach Hz-level.

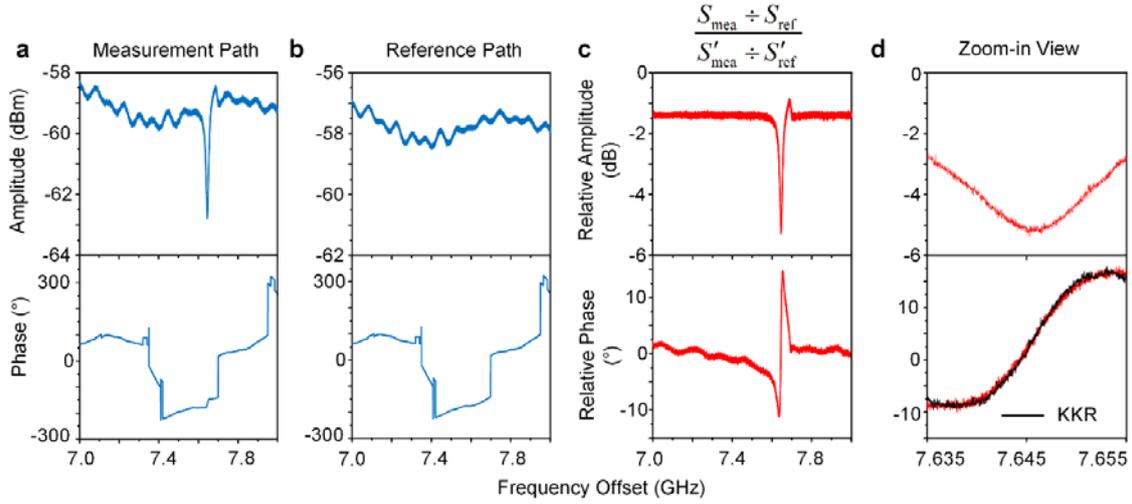

**Fig. 5** The magnitude and phase responses of the phase-shifted FBG. **a** The results in the measurement path, which contains the spectral response of the ODUT and an unknown time-varying signal introduced by environmental variations. **b** The results in the reference path, which only contains the unknown time-varying signal. **c** $S_{mea} \div S_{ref}$ can eliminate the unknown time-varying signal, $S'_{mea} \div S'_{ref}$ can get the spectral response of the system, $(S_{mea} \div S_{ref})/(S'_{mea} \div S'_{ref})$ is the final result after calibration. **d** The zoom-in view of the key region. KKR, Kramers-Kronig relationship. The red lines are the measurement results by the proposed OVA, and the black line is the phase response calculated via KKR.

**Implementation of ultra-large dynamic range**

Many high-Q optical devices not only require ultra-high resolution measurement, but also demand that the measurement can provide an ultra-large dynamic range. The maximum dynamic range of the OVA is determined by the ASR. In our implementation, the electrical part of the ASR has a noise floor of -118 dBm and a maximum input power of 22 dBm, and the two PDs (U$^2$T 2120RA) have a dark current of 5 nA (-119 dBm) and an output peak voltage of 325 mV (~0 dBm). Thus the power range after the PD is limited to [-118 dBm, 0 dBm] and the maximum dynamic range is restricted to 118 dB. In the conventional modulation-based OVA, the lower bound of the measurement is usually restricted by the residual sideband of the OSSB modulation, the high-order sidebands introduced by the modulation nonlinearity [23, 25], which would submerge the desired signal when the power of the desired signal is less than that of the high-order sidebands and raise the floor of the lower bound, and the residual signals in other unselected channels that can be suppressed to an extremely small value via the TOBPF and the square-law detection in the PD (see methods). For the proposed method, the AOM only produces ignorable residual sideband, and the high-order sidebands will not affect the measured results according to the above analysis. Therefore, the lower bound of the measurement is mainly determined by the residual signals in other channels, i.e., the SMSR after the mode selection module. As shown in Fig. 3, the SMSR of the selected carrier is varied from 21 to 42 dB. After square-law detection in the PD, the photocurrents beaten by the ±1st order sidebands and frequency-shifted carrier are given by $i_{PD+}=\eta E_{+1}(t)E^*_c(t)$ and $i_{PD-}=\eta E_c(t)E^*_{-1}(t)$, where $\eta$ =0.65 A/W is the responsivity of the PD, $E_{\pm1}(t)$ and $E_c(t)$ are the electric fields of the ±1st order sideband and the frequency-shifted carrier, respectively. The electrical power is $P= i^2R/2$ ($i$ is the peak current), where $R$= 50 Ω is the load impedance. Thus the power beaten by the ±1st order sidebands and the frequency-shifted carrier is $P_{PD+}=[\eta E_{+1}E^*_c]^2R/2$ and $P_{PD-}=[\eta E_c E^*_{-1}]^2R/2$, where $E_{\pm1}$ and $E_c$ are the complex amplitude of the ±1st order sideband and the frequency-shifted carrier, respectively. Given that both the frequency-shifted carriers and their ±1st order sidebands in the unselected channels are 21 to 42 dB lower than those in the selected channel, the power of the undesired beating signal would be 42-84 dB smaller than the useful beating signals due to the square-law detection in the PD, which indicates that the proposed method can be used to measure a notch response with a notch depth of 42-84 dB in the selected channel. However, the notch depth of an actual optical device is hard to exceed 70 dB, so we can choose an optical power from the ASG to balance the measurement of loss and gain. Assume the gain and loss only affect the sidebands, we get -118 dBm $\leqslant P_{PD\pm}=10\log_{10}(|\eta E_{\pm1}(t)E^*_c(t)|^2R)$ =$10\log_{10}([2\eta MHP_c]^2R/2) \leqslant 0$ dBm and $10\log_{10}(MHP_c)<13$ dBm (the maximum input power of the PD), where $M$ is the magnitude ratio of the sideband and the frequency-shift carrier, and $H$ is the magnitude response. In the experiment, the power of the frequency-shifted carrier $P_c$ is set to 0.74 dBm. When $20\log_{10}M$=-35.74 dB, i.e. the power of the sideband is -35 dBm, the proposed OVA can measure a -70-dB notch. In this case, a device with 48-dB peak gain would amplify the sideband to 13 dBm, i.e. the maximum input power of the PD. Therefore, the maximum dynamic range of the OVA is 118 dB, as shown in Fig. 6(a). The channel using the comb line in the marginal area would have a lower dynamic range, since it can merely measure a -42-dB notch and a 48-dB gain peak, reaching a 90 dB dynamic range.

In order to verify the ultra-large dynamic range of the proposed OVA, an experiment is performed with an ODUT consisting of two cascaded programmable optical filters (WaveShaper 4000s) and an 8-km single-mode fiber (SMF) pumped to provide stimulated Brillouin scattering (SBS) gain. The stop band of the filters is set to be around -60 dB and the 8-km SMF stimulates about 30-dB SBS gain. A comb line in the middle of the OFC is selected as the carrier. Fig. 6(b) shows the magnitude response of the ODUT measured by the proposed OVA (the red line) with a measurement range of 30 GHz. The peak gain provided by the SBS is 31.03 dB and the notch is 59.41 dB, indicating that the proposed method has a dynamic range of >90 dB. As a comparison, the result (the black line) achieved by an optical spectrum analyzer (OSA, APEX AP2041B) with a resolution of 5 MHz and a dynamic range of around 80 dB is also plotted. The two results agree well in most points except for the peak gain. Due to the higher measurement resolution and larger dynamic range, the proposed OVA can obtain a more accurate measurement. It should be noted that the dynamic range can be further extended if a variable optical attenuator is incorporated before the PD, which dynamically adjust the optical power to the PD.

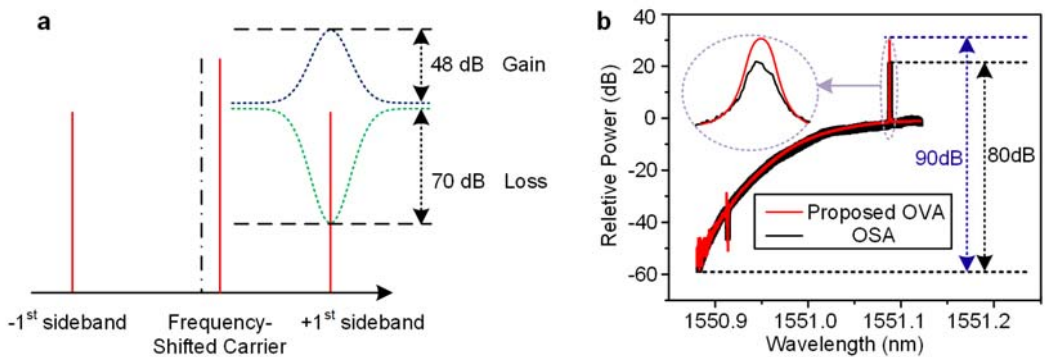

**Fig. 6** The dynamic range of the proposed OVA. **a** The illustration of the dynamic range. The dynamic range is the difference between the smallest and largest probe signal that can be detected by the ASR. **b** the magnitude response with >90-dB dynamic range measured by the proposed OVA (red line) and an OSA with 80-dB dynamic range (black line).

**Discussions**

We have presented a novel method to meet the measurement requirements of emerging optical devices for ultra-wide bandwidth, ultra-high resolution and ultra-large dynamic range simultaneously. The designed ASG and ASR extended the measurement range, removed the majority of measurement errors in the conventional modulation-based OVA, and therefore guaranteed a high resolution and a large dynamic range. By applying an OFC, OVA with a record resolution of 334 Hz (attometer level), a dynamic range of 90 dB and a measurement range of 1.025 THz was experimentally demonstrated. The proposed method has the potential to be a prevailing method for characterization of a variety of emerging optical devices and provide support for many frontier researches, such as on parity-time symmetry [1], optical nanoparticle detection [2], electromagnetically induced transparency [3], on-chip optical signal processing [4], ultra-sensitive optical sensing [5] and so on.

**Methods**

Mathematical proof is provided to explain the principle of the proposed OVA. The generated *N*-tone OFC signal writes:

$$E_{\text{OFC}} = \sum_{m=1}^{N} A_m \delta(\omega - \omega_1 - m\omega_{\text{rep}}), \tag{1}$$

where $\omega_1$ is the angular frequency of the first comb line, $\omega_{\text{rep}}$ is the frequency spacing of the OFC, and $A_m$ is the complex amplitude of the *m*th comb line. The selected *n*th comb line can be expressed as:

$$E'_{\text{OFC}} = A_n \delta(\omega - \omega_1 - n\omega_{\text{rep}}) + \sum_{\substack{m=1 \\ m \neq n}}^{N} \alpha_m A_m \delta(\omega - \omega_1 - m\omega_{\text{rep}}), 0 < \alpha_m < 1, \tag{2}$$

where $\alpha_m$ is the attenuation of the OBPF applied to the *m*th comb line, which is assumed to be a constant.

The output signal after the AOM is frequency-upshifted by $\Delta\omega$:

$$E_{\text{AOM}}(\omega) = bA_n \delta\left[\omega - (\omega_1 + n\omega_{\text{rep}} + \Delta\omega)\right] + b\sum_{\substack{m=1 \\ m \neq n}}^{N} \alpha_m A_m \delta\left[\omega - (\omega_1 + m\omega_{\text{rep}} + \Delta\omega)\right], \tag{3}$$

where *b* is the conversion efficiency of the AOM.

The other part of the optical signal is modulated by a sweeping RF signal $\omega_e$ at an MZM to generate a carrier-suppressed ODSB signal, which includes the sweeping ±1st-order sidebands and the unwanted components such as the residual carrier, the high-order sidebands, and the components in other channels. After the MZM, the output signal can be expressed as

$$E_{\text{MZM}}(\omega) = a_{+1}A_n\delta\left[\omega - (\omega_1 + n\omega_{\text{rep}} + \omega_e)\right] + a_{-1}A_n\delta\left[\omega - (\omega_1 + n\omega_{\text{rep}} - \omega_e)\right] + a_0 A_n\delta\left[\omega - (\omega_1 + n\omega_{\text{rep}})\right]$$
$$+ \sum_{\substack{i=-\infty \\ i\neq \pm 1,0}}^{\infty} a_i A_n\delta\left[\omega - (\omega_1 + n\omega_{\text{rep}} + i\omega_e)\right] + \sum_{\substack{m=1 \\ m\neq n}}^{N} \sum_{j=-\infty}^{\infty} \alpha_m a_j A_m \delta\left[\omega - (\omega_1 + m\omega_{\text{rep}} + j\omega_e)\right] \quad , \tag{4}$$

where $a_{\pm 1}$ and $a_0$ are the amplitudes of the sweeping ±1st order sidebands and the residual carrier, respectively.

The asymmetric optical probe signal is generated by combining Eqs. (3) and (4)

$$E_{\text{Probe}}(\omega) = bA_n\delta\left[\omega - (\omega_1 + n\omega_{\text{rep}} + \Delta\omega)\right] + b\sum_{\substack{m=1 \\ m\neq n}}^{N} \alpha_m A_m \delta\left[\omega - (\omega_1 + m\omega_{\text{rep}} + \Delta\omega)\right]$$
$$a_{+1}A_n\delta\left[\omega - (\omega_1 + n\omega_{\text{rep}} + \omega_e)\right] + a_{-1}A_n\delta\left[\omega - (\omega_1 + n\omega_{\text{rep}} - \omega_e)\right] + a_0 A_n\delta\left[\omega - (\omega_1 + n\omega_{\text{rep}})\right]. \tag{5}$$
$$+ \sum_{\substack{i=-\infty \\ i\neq \pm 1,0}}^{\infty} a_i A_n\delta\left[\omega - (\omega_1 + n\omega_{\text{rep}} + i\omega_e)\right] + \sum_{\substack{m=1 \\ m\neq n}}^{N} \sum_{j=-\infty}^{\infty} \alpha_m a_j A_m \delta\left[\omega - (\omega_1 + m\omega_{\text{rep}} + j\omega_e)\right]$$

Then, the asymmetric optical probe signal enters the measurement path and the reference path. In the measurement path, the asymmetric optical probe signal goes through the ODUT and carries on the magnitude and phase responses of the ODUT. The output signal can be written as:

$$E_{\text{mea}}(\omega) = E_{\text{Probe}}(\omega) \cdot H(\omega)$$
$$= bA_n H(\omega_1 + n\omega_{\text{rep}} + \Delta\omega)\delta\left[\omega - (\omega_1 + n\omega_{\text{rep}} + \Delta\omega)\right] + b\sum_{\substack{m=1 \\ m\neq n}}^{N} \alpha_m A_m H(\omega_1 + m\omega_{\text{rep}} + \Delta\omega)\delta\left[\omega - (\omega_1 + m\omega_{\text{rep}} + \Delta\omega)\right], \tag{6}$$
$$a_{+1}A_n H(\omega_1 + n\omega_{\text{rep}} + \omega_e)\delta\left[\omega - (\omega_1 + n\omega_{\text{rep}} + \omega_e)\right] + a_{-1}A_n H(\omega_1 + n\omega_{\text{rep}} - \omega_e)\delta\left[\omega - (\omega_1 + n\omega_{\text{rep}} - \omega_e)\right]$$
$$+ a_0 A_n H(\omega_1 + n\omega_{\text{rep}})\delta\left[\omega - (\omega_1 + n\omega_{\text{rep}})\right] + \sum_{\substack{i=-\infty \\ i\neq \pm 1,0}}^{\infty} a_i A_n H(\omega_1 + n\omega_{\text{rep}} + i\omega_e)\delta\left[\omega - (\omega_1 + n\omega_{\text{rep}} + i\omega_e)\right]$$
$$+ \sum_{\substack{m=1 \\ m\neq n}}^{N} \sum_{j=-\infty}^{\infty} \alpha_m a_j A_m H(\omega_1 + m\omega_{\text{rep}} + j\omega_e)\delta\left[\omega - (\omega_1 + m\omega_{\text{rep}} + j\omega_e)\right]$$

where $H(\omega) = H_{\text{sys}}(\omega)H_{\text{ODUT}}(\omega)$, $H_{\text{sys}}(\omega)$ and $H_{\text{ODUT}}(\omega)$ denote the transmission functions of the measurement system and the ODUT, respectively.

After square-law detection in a PD (PD1), the generated photocurrent contains components introduced by the sweeping sidebands, the frequency-shifted carrier, the residual carrier, the high order sidebands and the components in other wavelength channels. Due to the difference in frequency, the influence by the residual carrier and the high-order sidebands in all wavelength channels is removed. The components generated by the ±1st sidebands and the frequency-shifted carrier can be expressed as

$$
\begin{aligned}
i_{\text{mea},+1}(\omega_e - \Delta\omega) &= \eta a_{+1} A_n b H(\omega_1 + n\omega_{\text{rep}} + \omega_e) H^*(\omega_1 + n\omega_{\text{rep}} + \Delta\omega) \\
&+ \eta a_{+1} b \sum_{\substack{m=1 \\ m\neq n}}^{N} \left[ \alpha_m^2 A_m H(\omega_1 + m\omega_{\text{rep}} + \omega_e) H^*(\omega_1 + m\omega_{\text{rep}} + \Delta\omega) \right], \text{ if } \omega_e > \Delta\omega \\
i_{\text{mea},+1}(\Delta\omega - \omega_e) &= \eta a_{+1} A_n b H^*(\omega_1 + n\omega_{\text{rep}} + \omega_e) H(\omega_1 + n\omega_{\text{rep}} + \Delta\omega) \\
&+ \eta a_{-1} b \sum_{\substack{m=1 \\ m\neq n}}^{N} \left[ \alpha_m^2 A_m H^*(\omega_1 + m\omega_{\text{rep}} + \omega_e) H(\omega_1 + m\omega_{\text{rep}} + \Delta\omega) \right], \text{ if } \omega_e < \Delta\omega \\
i_{\text{mea},-1}(\omega_e + \Delta\omega) &= \eta a_{-1} A_n b H^*(\omega_1 + n\omega_{\text{rep}} - \omega_e) H(\omega_1 + n\omega_{\text{rep}} + \Delta\omega) \\
&+ \eta a_{-1} b \sum_{\substack{m=1 \\ m\neq n}}^{N} \left[ \alpha_m^2 A_m H^*(\omega_1 + m\omega_{\text{rep}} - \omega_e) H(\omega_1 + m\omega_{\text{rep}} + \Delta\omega) \right]
\end{aligned}
\quad , \quad (7)
$$

where $\eta$ is the responsivity of the PD. As can be seen in Eq. (7), the residual sidebands in other channels are suppressed by $\alpha_m^2$ times.

In the reference path, the other portion of the asymmetric optical probe signal is directly sent to a PD (PD2), generating a photocurrent including the same frequency components with Eq. (7). Without phase and magnitude changes due to the ODUT, we assume that the transmission function $H_{\text{ODUT}}(\omega)=1$ in the reference path. After the progress of $S_0 = S_{\text{mea}} \div S_{\text{ref}}$, time-varying measurement errors can be eliminated. To remove the time-invariant response $H_{\text{sys}}(\omega)$ of the measurement system, the two test ports can be directly connected to perform a calibration. In this case, the components with the frequencies of $\omega_e + \Delta\omega$, $|\omega_e - \Delta\omega|$ can be given by

$$
\begin{aligned}
i_{\text{cal},+1}(\omega_e - \Delta\omega) &= \eta a_{+1} A_n b H_{\text{sys}}(\omega_1 + n\omega_{\text{rep}} + \omega_e) H_{\text{sys}}^*(\omega_1 + n\omega_{\text{rep}} + \Delta\omega) \\
&+ \eta a_{+1} b \sum_{\substack{m=1 \\ m\neq n}}^{N} \left[ \alpha_m^2 A_m H_{\text{sys}}(\omega_1 + m\omega_{\text{rep}} + \omega_e) H_{\text{sys}}^*(\omega_1 + m\omega_{\text{rep}} + \Delta\omega) \right], \text{ if } \omega_e > \Delta\omega \\
i_{\text{cal},+1}(\Delta\omega - \omega_e) &= \eta a_{+1} A_n b H_{\text{sys}}^*(\omega_1 + n\omega_{\text{rep}} + \omega_e) H_{\text{sys}}(\omega_1 + n\omega_{\text{rep}} + \Delta\omega) \\
&+ \eta a_{-1} b \sum_{\substack{m=1 \\ m\neq n}}^{N} \left[ \alpha_m^2 A_m H_{\text{sys}}^*(\omega_1 + m\omega_{\text{rep}} + \omega_e) H_{\text{sys}}(\omega_1 + m\omega_{\text{rep}} + \Delta\omega) \right], \text{ if } \omega_e < \Delta\omega \\
i_{\text{cal},-1}(\omega_e + \Delta\omega) &= \eta a_{-1} A_n b H_{\text{sys}}^*(\omega_1 + n\omega_{\text{rep}} - \omega_e) H_{\text{sys}}(\omega_1 + n\omega_{\text{rep}} + \Delta\omega) \\
&+ \eta a_{-1} b \sum_{\substack{m=1 \\ m\neq n}}^{N} \left[ \alpha_m^2 A_m H_{\text{sys}}^*(\omega_1 + m\omega_{\text{rep}} - \omega_e) H_{\text{sys}}(\omega_1 + m\omega_{\text{rep}} + \Delta\omega) \right]
\end{aligned}
\quad . \quad (8)
$$

According to the fact that $α_m^2 ∈ $ [-84 dB,-42 dB], $α_m^2$ is an extremely small value, so the components in other wavelength channels can be ignored. Therefore, the transmission response of the ODUT can be achieved by Eqs. (7) and (8), which can be written as

$$H_{\text{ODUT}}(ω_1+mω_{\text{rep}}-ω_e) = \frac{i^*_{\text{mea},-1}(ω_e+Δω)}{i^*_{\text{cal},-1}(ω_e+Δω)H^*_{\text{ODUT}}(ω_1+nω_{\text{rep}}+Δω)}$$

$$H_{\text{ODUT}}(ω_1+mω_{\text{rep}}+ω_e) = \frac{i^*_{\text{mea},+1}(Δω-ω_e)}{i^*_{\text{cal},+1}(Δω-ω_e)H^*_{\text{ODUT}}(ω_1+nω_{\text{rep}}+Δω)}, \text{ if } ω_e < Δω, \quad (9)$$

$$H_{\text{ODUT}}(ω_1+mω_{\text{rep}}+ω_e) = \frac{i_{\text{mea},+1}(ω_e-Δω)}{i_{\text{cal},+1}(ω_e-Δω)H^*_{\text{ODUT}}(ω_1+nω_{\text{rep}}+Δω)}, \text{ if } ω_e > Δω$$

where $H_{\text{ODUT}}(ω_1+ω_{\text{rep}}+Δω_e)$ is a constant and represents the response at the wavelength of the frequency-shifted carrier. By scanning the RF frequency, the spectral responses of the ODUT $H_{\text{ODUT}}(ω_1+ω_{\text{rep}}-ω_e)$ and $H_{\text{ODUT}}(ω_1+ω_{\text{rep}}+ω_e)$ can be accurately achieved.

**Acknowledgements**

This work was supported in part by the National Natural Science Foundation of China (61527820); Jiangsu Provincial "333" Project (BRA2018042); Fundamental Research Funds for the Central Universities (NC2018005); Postgraduate Research & Practice Innovation Program of Jiangsu Province (KYCX18_0290); Project Funded by the Priority Academic Program Development of Jiangsu Higher Education Institutions.

**Author contributions**

T.Q. and S.L.P. conceived and initiated the project, T.Q., S.P.L. and B.D.G. designed and carried out the experiments, T. Q. and S.P.L. discussed and analyzed the theoretical analysis. T.Q, Z.Z.T. and S.L.P. contributed to the manuscript writing.

**Data availability**

The data supporting the findings of this study can be obtained from the corresponding author.